\documentclass[11pt]{article}
\usepackage{amsmath,amssymb,amsfonts,epsf}
\usepackage[nosort]{cite}
\usepackage[margin=1in]{geometry}

\usepackage{pslatex}

\def\nn{\nonumber}

\let\bm=\bibitem

\newcommand{\be}{\begin{equation}}
\newcommand{\ee}{\end{equation}}
\def\ba{\begin{array}}
\def\ea{\end{array}}

\def\sst#1{{\scriptscriptstyle #1}}

\def\dalemb#1#2{{\vbox{\hrule height .#2pt
        \hbox{\vrule width.#2pt height#1pt \kern#1pt
                \vrule width.#2pt}
        \hrule height.#2pt}}}

\newcommand{\bea}{\begin{eqnarray}}
\newcommand{\eea}{\end{eqnarray}}

\def\0{{\sst{(0)}}}
\def\1{{\sst{(1)}}}
\def\2{{\sst{(2)}}}
\def\3{{\sst{(3)}}}
\def\4{{\sst{(4)}}}
\def\5{{\sst{(5)}}}
\def\6{{\sst{(6)}}}
\def\7{{\sst{(7)}}}
\def\8{{\sst{(8)}}}

\begin{document}

\begin{center}\ \\ \vspace{0pt}
{\Large {\bf Deviations from Keplerian Orbits for Solar Sails}}\\ 
\vspace{10pt}
Roman Ya. Kezerashvili and Justin F. V\'azquez-Poritz

\vspace{5pt}
{\it Physics Department, New York City College of Technology, The City University of New York\\ 300 Jay Street, Brooklyn NY 11201, USA}

\vspace{5pt}
{\it The Graduate School and University Center, The City University of New York\\ 365 Fifth Avenue, New York NY 10016, USA}\\

\vspace{5pt}
{\tt rkezerashvili@citytech.cuny.edu}\ \ \ {\tt jvazquez-poritz@citytech.cuny.edu}
\end{center}

\vspace{10pt}
\noindent {\bf ABSTRACT}

\noindent It is shown that the curvature of spacetime, a possible net electric charge on the sun, a small positive cosmological constant and the oblateness of the sun, in conjunction with solar radiation pressure (SPR), affect the bound orbital motion of solar sails and lead to deviations from Kepler's third law for heliocentric and non-Keplerian orbits. With regards to the Lense-Thirring effect, the SRP increases the amount of precession per orbit for polar orbits. Non-Keplerian polar orbits exhibit an analog of the Lense-Thirring effect in which the orbital plane precesses around the sun. \\

\noindent {\bf Keywords:}\ \ solar sails, general relativity, Lense-Thirring effect\\ 

\pagestyle{empty}

\noindent {\bf INTRODUCTION}

The widely-discussed concept of the Solar Probe mission is an integral part
of the programs of many national space agencies. No previous spacecraft has traversed the region of less than 0.3 AU from the sun. However, during the last two
decades, a number of solar probes have been designed to perform in-situ near-sun measurements in a rather unusual
and difficult environment. With the success of the Ulysses mission and the
unprecedented opportunities offered by the SONO mission launched in 1995, a
golden era has dawned in solar physics. It is therefore not surprising that
a solar probe penetrating to a few solar radii has once again come
into focus \cite{guo}. The primary goals of the solar probe mission are to
explore the solar corona, unveil the formation and origin of the solar wind
and understand the mechanisms for storing and transporting energetic charged
particles. The solar probe will be a voyage of discovery to the center of
our solar system. 

The primary focus of the solar probe will be to address outstanding
questions regarding the structure, dynamics and physical
properties of the outer solar atmosphere. This region remains the last
unknown link between the solar output and the terrestrial environment. A
close-solar encounter concurrently provides the opportunity to conduct other
scientific investigations, such as the effects of general relativity on the
trajectory of a solar sail spacecraft. In contrast with the conventional solar
probe mission, a solar sail propelled (SSP) satellite allows for
non-Keplerian orbits, for which the orbital plane does not pass through the
center of mass of the sun and the SSP satellite is levitated above the sun.
General relativistic effects on a levitated near-sun solar sail is of particular interest.

One of the most basic laws that describes motion in the solar system is
Kepler's third law, \ which can be derived from Newton's law of gravitation
and provides a relationship between the period $T$ and the orbital radius $r$ for objects in circular orbit around the sun:

\begin{equation}
T^{2}=\frac{4\pi ^{2}}{GM}r^{3},  \label{Kepler}
\end{equation}%
where $G$ is the gravitational constant and $M$ is the mass of the sun. For elliptical orbits, we replace $r$ by the length $a$ of the semi-major axis of the ellipse. We will discuss deviations from Keplerian orbits due to phenomena within
conventional physics which can be observed from the motion of an SSP
satellite. In the general relativistic framework, objects tend to follow geodesics
on curved spacetime. At the same time, the sun emits electromagnetic radiation which produces an
external force on objects via the solar radiation pressure (SRP). Therefore, we can say that objects move in the \textit{photo-gravitational field} of the sun. It is of
interest to analyze how the trajectories of objects deviate from
geodesics due to the SRP. We will assume that the backreaction of the solar radiation on spacetime
is negligible. The purpose of this paper is to point out various sources of deviations from
Keplerian orbits and to discuss how the resulting change in period is enhanced
by the SRP to the degree in which it may be observed for some cases.\\

\noindent {\bf 1.\ \ \ GENERAL RELATIVISTIC EFFECTS FOR HELIOCENTRIC ORBITS}

We first consider the effect of the SRP on Keplerian orbits in the
Newtonian approximation of gravity. For simplicity, we start with the scenario in which the solar sail is in
circular orbit within the plane of the sun and the surface of the sail is directly facing the sun. Then both forces act along the same line and fall off with the heliocentric distance as $1/r^{2}$, where the SRP force is repulsive and the gravitational force is attractive. 
In this case, the mass of the sun in Kepler's third law (\ref{Kepler}) is effectively renormalized as
$M\rightarrow \tilde{M}=M-\kappa /G$, where $\kappa \equiv \frac{\eta L_{S}}{2\pi c\sigma }
$, $L_{S}=3.842\times 10^{26}$ W is the solar luminosity, and $\sigma$ is the mass per area of the sail-- a key design parameter for solar sails \cite{rk12}. The parameter $\eta$ lies in the range 
$0.5\leq \eta \leq 1$, where $\eta =0.5$ corresponds to the total absorption
of photons by the sail and $\eta =1$ corresponds to total reflection. We will consider an SSP satellite with the following specifications:
\bea\label{values}
r &=& 0.05\ \mbox{AU}\approx 7.48\times 10^9\ \mbox{m}\,,\nn\\
\eta &=& 0.85\,,\quad \sigma=0.00131\ \mbox{kg/m}^2\,.
\eea

We consider the simultaneous effects of the SRP and various general relativistic effects associated with the slowly rotating sun which has a mass $M$, angular momentum $J$, and possibly a small amount of net charge $Q$. The exterior spacetime in the vicinity of the sun is described by the large-distance limit of the Kerr-Newman metric:
\begin{equation}
ds^{2}=-fc^{2}dt^{2}-\frac{4GJ}{c^{2}r}\sin ^{2}\theta \ dtd\phi +\frac{%
dr^{2}}{f}+r^{2}(d\theta ^{2}+\sin ^{2}\theta \ d\phi ^{2})\,,\qquad f=1-\frac{2GM}{c^{2}r}+\frac{Gk_{e}Q^{2}}{c^{4}r^{2}}\,.
\label{Stat+Kerr+Q}
\end{equation}
where the Coulomb constant $k_{e}=8.988\times 10^{9}$ N m$^{2}$/C$^{2}$. We first consider the case of $Q=0$, for which (\ref{Stat+Kerr+Q}) reduces to the Kerr metric.
The constants of motion $E\equiv -p_{t}/m$, $L_{z}\equiv p_{\phi }/m$ and
$p^{\theta }=\frac{m}{r^{2}}\sqrt{P-\frac{L_{z}^{2}}{\tan ^{2}\theta }}$ are associated with the energy per mass, the component of angular momentum per mass that is normal to the equatorial plane, and the total angular momentum per mass, respectively. $P=0$
corresponds to motion in the equatorial plane, for which $\theta =\pi /2$
and $p^{\theta }=0$. No other orbits can lie within a fixed plane. In the
presence of the SRP, the orbital equation is \cite{poritz}
\begin{equation}
\left( \frac{dr}{d\tau }\right) ^{2}=\frac{E^{2}}{c^{2}}-c^{2}+\frac{2G%
\tilde{M}}{r}-\frac{f(L_{z}^{2}+P)}{r^{2}}-\frac{4GJEL_{z}}{c^{4}r^{3}}\,.
\label{Kerr}
\end{equation}
From (\ref{Kerr}), we find that circular orbits $\left(\frac{dr}{d\tau }=\frac{d^{2}r}{d\tau ^{2}}=0\right)$ in the equatorial plane have an orbital period $T$ of
\begin{equation}
T^{2}\approx \frac{4\pi ^{2}r^{3}}{G\tilde{M}}\left[ 1+\kappa \frac{%
(c^{2}r-4GM)}{(c^{2}r-2GM)^{2}}\right] \left[ 1+\frac{2\sqrt{G}J\Big(1+\frac{%
4\kappa (c^{2}r-GM)}{(c^{2}r-2GM)^{2}}\Big)}{c^{2}\sqrt{\tilde{M}}r^{3/2}%
\sqrt{1+\frac{\kappa (c^{2}r-4GM)}{(c^{2}r-2GM)^{2}}}}\right] \,.
\label{Tkerr}
\end{equation}
We have presented $T^{2}$ as a product of three factors. The first factor is
the same as in the Newtonian approximation for gravity in the presence of
the SRP, and embodies the fact that the SRP effectively renormalizes the
solar mass. The second factor shows the deviation from Keplerian orbits due
to the simultaneous effects of the SRP and the curvature of spacetime outside of
a static central body. The third factor gives the deviation from Keplerian
orbits from the combined effects of the SRP and frame dragging due to the
rotation of the sun. Note that the second and third factors in (\ref{Tkerr})
involve both the solar mass $M$ and the parameter $\kappa$ separately,
rather than simply in the combination of the renormalized solar mass $\tilde{%
M}$. For a static spacetime ($J=0$) and for no SRP effects ($\kappa =0$), (\ref{Tkerr}) reduces respectively to 
\begin{equation}
T^{2}=\frac{4\pi ^{2}r^{3}}{G\tilde{M}}%
\left[ 1+\kappa \ \frac{c^{2}r-4GM}{(c^{2}r-2GM)^{2}}\right]\,,\qquad
T^{2}=\frac{4\pi ^{2}r^{3}}{GM}\left[ 1+\frac{2\sqrt{G}J}{c^{2}\sqrt{M}%
r^{3/2}}\right] \,.
\label{period}
\end{equation}%
%
%
The first expression in (\ref{period}) shows that it is only the simultaneous effects of the static curvature of spacetime and the SRP that lead to a radial dependent (and thus energy dependent) deviation from Kepler's third law. For the specification given in (\ref{values}), we find that this yields an increase in the period of about $0.6$ s. The second expression in (\ref{period}) shows that, in the absence of the SRP, there is still a deviation from Keplerian orbits due to frame dragging.
%
%
However, the SRP enhances the effect of frame dragging, since it is the renormalized mass $\tilde{M}$ that appears in the last factor. The speed of the outer layer of the sun at its equator
is $v\approx 2000$ m/s at the equatorial radius $R\approx 7\times 10^{8}$ m,
and we will assume that the core of the sun rotates with the same angular
speed. Therefore, $J={\textstyle{\frac{\scriptstyle2}{\scriptstyle5}}}%
MvR\approx 10^{42}$ kg m$^{2}$/s. Without the effects of the SRP, frame
dragging leads to an increase (decrease) in the period of only $4\times
10^{-5}$ s for a prograde (retrograde) orbit. With the SRP and the
specifications in (\ref{values}), frame dragging leads to a change in
period of about $0.01$ s.

The orbital plane of a polar orbit which passes through the poles of the sun will
precess in the $\phi $ direction, which is a well-known effect of frame
dragging called the Lense-Thirring effect. Following \cite{poritzkez}, we can
find the approximate angle of precession for a polar orbit during one
orbital period up to linear order in $J$:
\begin{equation}
\Delta \phi \approx \frac{4\pi GJ}{c^{2}\sqrt{G\tilde{M}r^{3}}}\,,
\end{equation}%
%
%
%
%
%
which is increased by the SRP due to the corresponding increase in period.
For both a standard and an SSP satellite in a polar orbit at $r=0.05$ AU, we
find the rate of precession to be about $0.03$ arcseconds per year.

We will now consider the effect that a small amount of net charge $Q$ on the
sun would have on an SSP satellite with charge $q$. It has been suggested
that the sun has a net charge of up to $Q\approx 77$ C \cite{solarcharge}.
Taking into account the electric force, we find that 
\begin{equation}
T^{2}\approx \frac{4\pi ^{2}r^{3}}{G\tilde{M}}\Big[1+\frac{\kappa }{c^{2}r}+%
\frac{k_{e}qQ}{Gm\tilde{M}}+\frac{k_{e}Q^{2}}{c^{2}\tilde{M}r}+\frac{%
k_{e}^{2}q^{2}Q^{2}}{G^{2}m^{2}\tilde{M}^{2}}\Big]\,,
\end{equation}%
%
%
where we have kept only terms up to order $Q^{2}$. Note the $Q^{2}$ term in
the period, due to the backreaction from the sun's charge on the geometry,
is present even for a neutral SSP satellite. However, this term increases
the period by an amount of only $10^{-35}$ s ($10^{-38}$ s without the SRP),
which reflects the fact that the backreaction of the charge on the geometry
is negligible, as to be expected.

Primarily due to the photoelectric effect, but also the Compton effect and
electron-positron pair production, a solar sail made from Beryllium, for
example, will equilibrate to a charge per area of $0.065$ C/m$^2$ 
\cite{rk2008}. For an SSP satellite with the specifications given in \cite{poritz,poritzkez} and a mass of $1000$ kg, this gives a charge $q=5\times
10^4$ C and an increase in period of about $230$ s ($0.05$ s without the
SRP). Thus, due to the SRP, even a small charge $Q$ could certainly yield a
measurable increase in the period, making this a potentially powerful test
for net charge on the sun.

Since the SRP enhances a variety of small effects, including those
associated with spacetime curvature, one could ask whether an SSP satellite
could be used to test for the presence of a cosmological constant. In fact,
supernovae observations suggest that our universe might have a very small
positive cosmological constant \cite{cosm1,cosm2}. In the presence of a
cosmological constant $\Lambda$ and neglecting the effects of the sun's rotation as well as possible net charge, the spacetime in the vicinity of the sun
is described by a metric of the form (\ref{Stat+Kerr+Q}), where the function $%
f$ is now given by 
$f=1-\frac{2GM}{c^{2}r}-\frac{\Lambda r^{2}}{3}$. This yields
\begin{equation}
T^{2}\approx \frac{4\pi ^{2}r^{3}}{G\tilde{M}}\Big[1+\frac{c^{2}\Lambda }{3G%
\tilde{M}}r^{3}\Big]\,.
\end{equation}%
%
%
For $\Lambda \approx 10^{-52}$ m$^{-2}$ along with the specifications in
(\ref{values}), this leads to an insignificant increase in the orbital period of $%
10^{-17}$ s ($10^{-21}$ s without the SRP). While one might be able specify
values of $\sigma $ and $r$ such that the correction factor is actually
significant, the period would then be too large to make this a feasible test
for the presence of a cosmological constant.\\

\noindent {\bf 2.\ \ \ GENERAL RELATIVISTIC EFFECTS FOR ORBITS OUT OF THE PLANE OF THE SUN}

\begin{figure}[th]
\begin{center}
$\begin{array}{c@{\hspace{0.7in}}c}
\epsfxsize=2.0in \epsffile{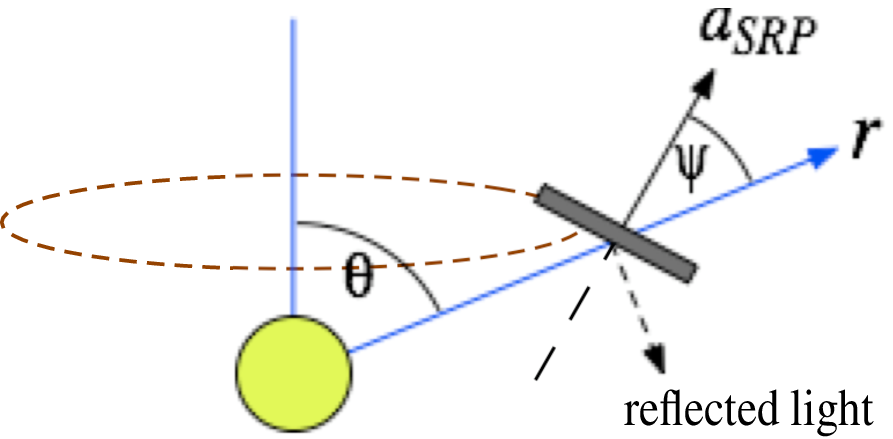} &
\epsfxsize=1.8in \epsffile{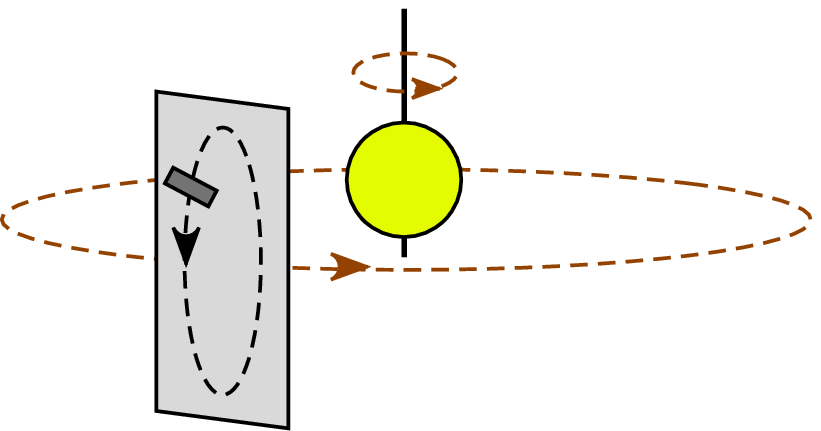}
\end{array}$
\end{center}
\caption[FIG. \arabic{figure}.]{\footnotesize{Orienting a solar
sail at an appropriate pitch angle $\protect\psi $ renders it possible to have a
circular orbit outside of the plane of the sun, as shown in the left plot. The Lense-Thirring effect for non-Keplerian polar orbits is shown in the right plot.}}
   \label{fig4}
\end{figure}

We will briefly review non-Keplerian orbits within the framework of
Newtonian gravity. The plane of a non-Keplerian orbit does not pass through
the center of mass of the sun, and the SSP satellite is levitated above the
sun, as shown in Figure \ref{fig4}. We orient our spherical coordinate system
so that the SSP satellite orbits at constant heliocentric distance $r$ and polar angle $\theta $. 
A non-Keplerian orbit can be maintained with a suitable pitch angle $\psi$
relative to the radially outgoing solar radiation hitting the surface of the
sail. This enables one to control the direction of thrust due to the
reflected portion of sunlight. By changing the pitch angle, one can also
transfer the SSP satellite between different non-Keplerian orbits whose
orbital planes have different orientations, as well as between non-Keplerian
orbits and orbits within the plane of the sun.


Recall that $\eta$ parameterizes the portion of light that is reflected by the sail. The reflected light pushes the solar sail at an angle $\psi $ relative to the
radial direction, while the absorbed light pushes the solar sail radially outwards  \cite{nonkepler1, rk1}. 
In the Newtonian approximation for
circular non-Keplerian orbits \cite{poritzkez},
\begin{equation}
T^{2}=\frac{4\pi ^{2}}{G\tilde{M}^{\ast}}\ r^{3}\sin ^{2}\theta \,,\qquad  
\tilde{M}^{\ast}\equiv M-\frac{L_S}{2\pi Gc\sigma}\left[ (1-\eta )+(2\eta -1)\cos
^{2}\psi \right]\,. 
\label{T1}
\end{equation}%
%
%
%
%
%
%

Now we consider the effects of curved spacetime on non-Keplerian orbits. 
Keeping only the leading correction due to spacetime curvature for orbits parallel to the equatorial plane of the sun, the period can be found to be
%
%
%
%
%
\begin{equation}
T^{2}\approx \frac{4\pi ^{2}}{G\tilde{M}^{\ast }}\sin ^{2}\theta \ r^{3}\Big[%
1+\frac{ G(M-\tilde M^{\ast})}{%
c^{2}r}\Big]\left[ 1+\frac{2\sqrt{G}J\sin \theta }{c^{2}\sqrt{\tilde{M}%
^{\ast }}r^{3/2}}\right] \,.  \label{nonkeplerTkerr}
\end{equation}%
In (\ref{nonkeplerTkerr}) $T^{2}$ is a product of three factors. The first
factor is the same as for non-Keplerian orbits in the Newtonian
approximation for gravity. The second factor is due to the simultaneous
effects of the SRP and the static curvature of spacetime. The third factor embodies the combined effects of the SRP and
frame dragging due to the rotation of the sun. 
As was the case for orbits within the plane of the sun, the SRP increases
the effect that frame dragging has on the orbital period, since it is the
renormalized mass $\tilde{M}^{\ast }$ that appears in the last factor.

There is an analog of the Lense-Thirring effect for non-Keplerian orbits. Namely, frame dragging causes the plane of non-Keplerian orbits parallel to polar orbits to precess around the sun, as shown in
the right diagram of Figure \ref{fig4}. Although we do not present the resulting equations in their entirety since they are rather cumbersome, the angle of precession can be approximated by 
\begin{equation}
\Delta \phi \approx \frac{4\pi GJ\sin \theta }{c^{2}\sqrt{G\tilde{M}^{\ast
}r^{3}}}\,.
\end{equation}

\noindent {\bf 3.\ \ \ THE OBLATENESS OF THE SUN}

We now consider the deviation from Kepler's third law on non-Keplerian circular orbits due to the oblateness of the sun. The oblateness of the sun has recently been measured with unprecedented
precision to be as much as $J_{2}\approx 9\times 10^{-6}$ during active
phases of the solar cycle \cite{oblatesun}.
Working in Newtonian gravity, the external
gravitational potential of an oblate spheroid is given by \cite{oblateness} 
\begin{equation}
V=-\frac{G\tilde{M}}{r}+\frac{GM}{r}\sum_{n=2}^{\infty }J_{n}\Big(\frac{R}{r}%
\Big)^{n}P_{n}(\cos \theta )\,,
\end{equation}%
%
%
%
%
%
%
%
%
%
where $J_{n}$ are the multipole mass moments, $R$ is the equatorial radius
of the sun and $P_{n}$ are the Legendre polynomials. The effective
mass in the first term is renormalized by the solar radiation pressure,
whereas the multipole mass moments are not affected. A consideration of the first sub-leading term in $J_2$ due to the oblateness of the sun leads to the relation
\begin{equation}
T^{2}\approx \frac{4\pi ^{2}r^{3}\sin ^{2}\theta }{G\tilde{M}^{\ast }}\left(
1+\frac{3MJ_{2}R^{2}}{2\tilde{M}^{\ast }r^{2}}(1-3\cos ^{2}\theta )\right)
\,.  \label{OblatTnonK}
\end{equation}%
Interestingly enough, the oblateness of the sun increases the period for
non-Keplerian orbits within the angular range of approximately $55^{\circ }\le \theta
< 90^{\circ }$, whereas the period is decreased for $\theta <55^{\circ }$. The case for which the circular orbit of the SSP satelite is confined to the equatorial plane can be obtained from (\ref{OblatTnonK}) by putting $\theta=90^{\circ}$. As an example, without the SRP, the oblateness of the sun increases the period of orbits within the plane of the sun by about $0.02$ s. With the SRP and the
specifications given in (\ref{values}), the sun's oblateness increases the period by about $105$ s.

For non-circular orbits, the perihelion shift per period is given by 
\begin{equation}
\Delta \phi \approx -\ \frac{3\pi MJ_{2}R^{2}}{\tilde{M}r^{2}}\,.
\end{equation}%
%
%
%
The characteristics of the precession due to the oblateness of the sun are
rather different than general relativistic precession. For instance,
precession due to oblateness occurs in the reverse direction relative to the orbit. 
For a conventional satellite at $r=0.05$ AU, the
precession rate is about $14$ arcseconds per year. For our SPP
satellite, the precession rate increases by a factor of $\sqrt{M/\tilde{M}}$
to about $235$ arcseconds per year. Although the SRP
increases the rate of precession due to the oblateness of the sun, it can be shown that the precession rate due to spacetime curvature is decreased by the SRP. Perhaps the SRP could be used to disentangle these two types of precession.\\


\noindent {\bf CONCLUSIONS}

We have considered the effects of various phenomena along with the SRP on
bound orbits of a solar sail propelled satellite. The consideration of the
SRP on its own leads to a renormalization of the effective solar mass.
However, when the SRP is coupled to other effects such as spacetime
curvature, the orbital characteristics depend on the solar
mass $M$ and the solar sail parameter $\kappa $ separately, and not simply
on the combination of a renormalized mass. The SRP affects the period of the
SSP satellite in two ways: by effectively decreasing the solar mass, thereby
increasing the period, and by enhancing the effects of other phenomena by
three orders of magnitude or more, rendering some of them detectable.  When
the SRP is not taken into account, the static curvature of spacetime does
not alter the period of circular orbits as given by Newtonian gravity. However, when the
effects of spacetime curvature and the SRP are
considered simultaneously for an SSP satellite, there is a
deviation from Kepler's third law.

There are two sources of perihelion shift for non-circular orbits. Spacetime curvature causes the perihelion to be shifted in the direction of the orbit, whereas the oblateness of the sun causes the perihelion to be shifted in the reverse direction. Without the SRP, the
perihelion shift due to spacetime curvature is generally larger than that
due to the oblateness of the sun. However, this is not the case when the
effects of the SRP are taken into account on an SSP satellite. 

For orbits within the plane of the sun as well as non-Keplerian orbits,
we have considered the Lense-Thirring effect, which is the precession of
polar orbits due to the rotation of the sun. The rate of precession is not altered by the
SRP up to leading order in the angular momentum of the sun. In particular, any satellite in a polar orbit at a distance of $0.05$ AU from the sun will experience a rate of precession of about $0.03$ arcseconds per year.

Observations of the trajectory of a near-solar mission would provide an interesting confirmation of these various phenomena.


\end{document}